# Ferromagnetic Instability in $AFe_4Sb_{12}$ (A = Ca, Sr, and Ba)

E. Matsuoka[1], K. Hayashi[1], A. Ikeda[1], K. Tanaka[1], and T. Takabatake[1,2]
[1]*Department of Quantum Matter, ADSM, and* [2]*Materials Science Center, N-BARD, Hiroshima University, Kagamiyama1-3-1, Higashi-Hiroshima 739-8530, Japan*

W. Higemoto
*Advanced Science Research Center, Japan Atomic Energy Research Institute, Tokai, Ibaraki 319-1195, Japan*

M. Matsumura
*Faculty of Science, Kochi University, Kochi 780-8520, Japan*

Magnetic, transport and thermal properties of $AFe_4Sb_{12}$ (A = Ca, Sr, Ba) are reported. All three compounds show a maximum in both the magnetic susceptibility and thermopower at 50 K, and a large electronic specific heat coefficient of 100 mJ/mol $K^2$. These properties are the characteristics of a nearly ferromagnetic metal. Furthermore, a remanent moment of the order of $10^{-3}\mu_B$/Fe was observed below 54, 48, and 40 K for A = Ca, Sr, and Ba, respectively. The volume fraction of the ferromagnetic component was estimated to be 10-20% by zero-field μSR measurements.

The concept of itinerant magnetism has been developed in the past three decades [1]. In weakly or nearly ferromagnetic systems, the temperature dependence of the amplitude of local spin density results in the Curie-Weiss behavior of magnetic susceptibility at high temperatures. In an itinerant ferromagnet, the ordered moment can be much smaller than the effective moment, and the magnetic ground state is very sensitive to atomic disorder, lattice deformation, and structural defects, as found for example in $Ni_3Al$ and $CaRuO_3$ [2,3]. Low-energy spin fluctuations significantly influence the transport and thermal properties at low temperatures. For example, spin fluctuation scattering gives rise to quadratic $T$ dependence of the electrical resistivity, and the temperature variation of fluctuations leads to the significant enhancement of the electronic specific heat. Furthermore, spin fluctuations in $ZrZn_2$ and $Sr_2RuO_4$ are believed to act as glue for electron pairing in unconventional superconducting states [4,5].

$NaFe_4Sb_{12}$ is a newly discovered weak itinerant ferromagnet with a rather high $T_c$ of 85 K and a small ordered moment of 0.25 $\mu_B$/Fe atom [6]. This compound belongs to the family of "filled skutterudites" with a chemical formula of $MT_4X_{12}$, where M is an alkali, alkaline-earth or a rare-earth metal, T is Fe, Ru or Os, and X is P, As, or Sb. The T atom is located at the center of an $X_6$ octahedron, and the cornersharing $TX_6$ octahedra form a tilted array. Among this family, rare-earth filled compounds have been extensively studied because of their novel physical properties such as non-Fermi-liquid behavior in $CeRu_4Sb_{12}$ [7] and unconventional superconductivity in $PrOs_4Sb_{12}$ [8]. Band structure calculations [9] have suggested that these phenomena are caused by strong mixing effect of 4*f* electrons of Ce or Pr and Sb 5*p* electrons. However, the role of *d* electrons of T atoms in the physical properties has not been well understood so far. The discovery of itinerant ferromagnetism in NaFe$_4$Sb$_{12}$ with monovalent Na ion has revealed the strong dependence of the *d* electron magnetism on the valence of the filler ion. In fact, LaFe$_4$Sb$_{12}$ with trivalent La ion stays in an enhanced paramagnetic state [10]. These contrasting magnetic behaviors between compounds with monovalent and trivalent filler ions in turn suggest that the ground state of alkaline-earth filled skutterudites AFe$_4$Sb$_{12}$ (A = Ca, Sr, Ba) with divalent A ions should be close to a magnetic critical point between an enhanced paramagnetic state and an itinerant ferromagnetic state. In this Letter, we report the magnetic, transport, and thermal properties of AFe$_4$Sb$_{12}$ compounds with A = Ca, Sr, and Ba. The results show that all three are in a nearly ferromagnetic state with strong spin fluctuations of Fe 3*d* electrons. Furthermore, weak and inhomogeneous ferromagnetism has been observed below approximately 50 K by the magnetization and muon spin relaxation (μSR) measurements.

The first investigation of magnetic properties for AFe$_4$Sb$_{12}$ (A = Ca, Sr, Ba) was done by Danebrock *et al*. [11]. The magnetic susceptibility above 100 K followed a Curie-Weiss law with an effective magnetic moment of 2 μ$_B$/Fe and a Weiss temperature ranging from –36 K for A = Ba to 3 K for A = Ca. This result was explained by a localized moment picture assuming that two Fe$^{3+}$/f.u. carry a moment of 2.8μ$_B$/Fe and two Fe$^{2+}$/f.u. no moment. More recently, A. Leithe-Jasper *et al*., [12] however, have interpreted similar results of magnetic susceptibility measurements for A = Ca and Sr as the characteristic of a nearly ferromagnetic system. No systematic measurements of transport properties have been reported, probably because of the difficulty in synthesizing dense samples. It was only noted that the electrical resistivities of $CaFe_4Sb_{12}$ and $BaFe_4Sb_{12}$ show an S-shape temperature dependence [6].

Our samples of $AFe_4Sb_{12}$ were synthesized by following the method reported by Danebrock *et al*. [11]. Starting from distilled dendritic pieces of Ca (99.9%), Sr (99.9%), Ba (99.9%), Fe powder (99.99%), and Sb lumps (99.9999%), we prepared binary antimonides, $CaSb_2$, $SrSb_2$, $BaSb_3$, and $FeSb_2$ in an evacuated quartz tube. The antimonides were mixed with Sb in a composition of A: T: Sb = 2 : 4 : 14 and pressed into a pellet in a glove box. The pellet was sealed in an evacuated quartz tube, heated to 973 K, and kept at this temperature for 24 hours. X-ray powder diffraction analysis of the reactant revealed the presence of $ASb_2$ and Sb in addition to the main phase of the filled skutterudite. These impurities were totally removed by soaking ground powder in aqua regia. In fact, all of the diffraction peaks were indexed as the $LaFe_4P_{12}$-type cubic structure. The purified powder samples are hereafter called powder sample. The lattice parameters were determined to be 9.156, 9.177, and 9.199 Å for A = Ca, Sr, and Ba, respectively, which are in good agreement with the reported values [13]. Reference compounds $LaFe_4Sb_{12}$ and $BaRu_4Sb_{12}$ were also synthesized by a similar manner as mentioned above.

For transport property measurements, high density pellets were made by a spark-plasma sintering (SPS) technique using the DR. SINTER LAB (Sumitomo Coal Mining). The typical density of samples sintered at 743 K under a pressure of 400 MPa for 30 min. was 94% of the theoretical value. The samples thus prepared are hereafter called SPS samples. The chemical composition was determined by electron-probe microanalysis (EPMA) with a wavelength dispersive JEOL JXA-8200 system. The analysis for 30 sampling spots gave the average compositions; A: Fe: Sb; 0.967 : 4.063 : 11.97 for A = Ca, 1.019 : 4.044 : 11.94 for A = Sr and 0.989 : 4.040 : 11.97 for A = Ba. This confirms the full occupancy of A ions in these samples. For the La compound, however, the La occupancy was found to be 0.86. It is noteworthy that the x-ray diffraction peaks of the SPS samples are broadened when compared with those of powder samples. As an example, the profiles of a peak indexed as (763) are compared in the inset of Fig. 1(b). One notices that the double peaks due to the Cu $K\alpha 1$ and $K\alpha 2$ radiations are well separated for the powder sample, while the two peaks merge for the SPS sample, keeping the center unchanged. This fact indicates that the samples undergo non-uniform distortion by the SPS process. The peak width did not change by the annealing at 743 K, above which the $AFe_4Sb_{12}$ phase decomposes.

The electrical resistivity was measured by a DC four-probe method from 1.3 to 300 K. The thermopower measurements from 4.2 to 300 K and from 80 to 500 K were performed by using a home-made setup with a differential method and a commercial MMR Seebeck effect measurement system, respectively. The measurement of the specific heat was carried out by a relaxation method (Quantum Design PPMS) from 2 to 300 K. The magnetization was measured by a SQUID magnetometer (Quantum Design MPMS) in a temperature range between 2 and 370 K.

Figs. 1(a) and 1(b) show, respectively, the temperature variations of the electrical resistivity $\rho(T)$ and thermopower $S(T)$ for AFe$_4$Sb$_{12}$, LaFe$_4$Sb$_{12}$, and BaRu$_4$Sb$_{12}$. The high density of the SPS samples manifests itself in the rather low electrical resistivity of 350-450 $\mu\Omega$ cm at 300 K, which is far less than the reported value [6]. In contrast to a monotonic decrease in $\rho(T)$ for BaRu$_4$Sb$_{12}$, $\rho(T)$ of AFe$_4$Sb$_{12}$ pass through a shoulder at 70 K and sharply decrease on cooling. This behavior resembles that of a nearly ferromagnetic metal like YCo$_2$ [14]. In such a system, $\rho(T)$ can be described by adding the spin fluctuation scattering term $\rho_{sf}(T)$ to the normal metallic resistivity as follows; $\rho(T) = \rho_0 + \rho_{ph}(T) + \rho_{sf}(T)$, where the first and second terms are residual and phonon scattering resistivity, respectively [14]. An expression of $\rho_{sf}(T) = AT^2$ with an enhanced coefficient $A$ at temperatures below the characteristic temperature $T_{sf}$ was derived by the self-consistent renormalization (SCR) theory for spin fluctuations [15]. The temperature at which d$\rho$/d$T$ shows a maximum gives an estimation of $T_{sf}$ ~50 K [14]. As is shown in the inset, $\rho(T)$ for AFe$_4$Sb$_{12}$ follow the $T^2$ dependence below about 9 K with $A = 7.76 \times 10^{-3}$, $1.34 \times 10^{-2}$, and $1.66 \times 10^{-2}$ $\mu\Omega$cm K$^{-2}$ for A = Ca, Sr, and Ba, respectively. We will discuss later the relation between the $A$ value and the electronic specific coefficient. The bend in $\rho(T)$ around 70 K is a result of the saturation of $\rho_{sf}(T)$ above $T_{sf}$.

The $S(T)$ for AFe$_4$Sb$_{12}$ is characterized by a local maximum at about 50 K, which coincides with the above determined value of $T_{sf}$. This agreement indicates that the local maximum in $S(T)$ is a result of saturation of the contribution from spin fluctuations upon elevating temperature above $T_{sf}$. For the nearly ferromagnetic metal YCo$_2$, a local minimum in $S(T)$ was observed near the $T_{sf}$ [14]. It was attributed to a paramagnon-drag effect, in which the thermally non-equilibrium paramagnons drag charge-carriers along the temperature gradient. For LaFe$_4$Sb$_{12}$, a change in sign from positive to negative at 80 K hints to the presence of more than one type of charge carrier as was reported [16]. For BaRu$_4$Sb$_{12}$, the rather small value of $S(T)$ together with the absence of a shoulder in $\rho(T)$ is consistent with the diamagnetic behavior in the magnetic susceptibility (not shown).

The specific heat $C(T)$ measured on SPS samples is presented in Fig. 2. For the three compounds, $C/T$ data vs. $T^2$ plotted in the inset is linear below 6 K, and the extrapolation to $T=0$ yields the electronic specific heat coefficient $\gamma$ to be 118, 87, and 104 mJ/mol K$^2$ for A = Ca, Sr, and Ba, respectively. These values are one order of magnitude larger than 16 and 10 mJ/mol K$^2$, observed for Ru compounds with A = Sr and Ba, respectively. The enhancement of the $\gamma$ value in the Fe compounds reflects the large electronic density of states near the Fermi level. In a nearly ferromagnetic system, the $\gamma$ value is related to the coefficient $A$ of the $T^2$ term in $\rho(T)$. The ratio $A/\gamma_{Fe}^2$, where $\gamma_{Fe}$ is $\gamma$ per one Fe atom, is $8.9 \times 10^{-6}$, $2.8 \times 10^{-5}$, and $2.5 \times 10^{-5}$ ($\mu\Omega$ cm/ K$^2$)/(mJ/Fe-mol K$^2$)$^2$ for A = Ca, Sr, and Ba, respectively. These values are close to the so-called Kadowaki-Woods value ($1.0 \times 10^{-5}$ $\mu\Omega$ cm K$^{-2}$/(mJ/mol K$^2$)$^2$) [17]. Here, it is noteworthy that a band structure calculation of AFe$_4$Sb$_{12}$ has shown that the narrow Fe 3$d$ band which is hybridized with the Sb 5$p$ electron state is located just below the Fermi level, and the condition for a ferromagnetic ground state is satisfied [18].

The inverse magnetic susceptibility $B/M$ ($B = 5$ T) for powder samples of AFe$_4$Sb$_{12}$ and LaFe$_4$Sb$_{12}$ is presented in Fig. 3. The data for SPS samples measured at 5 T were almost the same. The high-temperature data above 200 K can be described by the Curie-Weiss law with the effective magnetic moment $\mu_{eff}$ of 1.52, 1.47, and 1.50$\mu_B$/Fe and Weiss temperature $\theta_P$ of 54, 53, and 31 K for A = Ca, Sr and Ba, respectively. The $\theta_P$'s are one order of magnitude larger than that of LaFe$_4$Sb$_{12}$ (3.4 K). The largely positive $\theta_P$'s indicate that the ferromagnetic interaction between Fe 3$d$ electrons in AFe$_4$Sb$_{12}$ is stronger than in LaFe$_4$Sb$_{12}$. As shown in the inset (a) of Fig. 3, the isothermal magnetization curves $M(B)$ of AFe$_4$Sb$_{12}$ at 2 K increase almost linearly with increasing field up to 5 T, where the largest value of $M(B)$ among the three is only 0.08 $\mu_B$/Fe for A = Ca. This extremely small moment compared with $\mu_{eff}$ means that the Curie-Weiss behavior is not due to the localized 3$d$ moments as proposed by Danebrock et al. [11] but due to itinerant magnetism of Fe 3$d$ electrons as argued by Laithe-Jasper et al. [12]. One characteristic of a nearly ferromagnetic system is a broad maximum in the magnetic susceptibility near the $T_{sf}$. In the inset (b) of Fig.3, we notice a maximum in $M/B$ for A = Sr and Ba at about 50 K, which agrees with $T_{sf}$ estimated from the temperature variations of $\rho(T)$ and $S(T)$. The nature of the nearly ferromagnetic state has been corroborated by $^{123}$Sb-NQR experiment for powder sample of SrFe$_4$Sb$_{12}$. It was found that the spin-lattice relaxation time $T_1$ follows the Curie-Weiss behavior $(T_1T)^{-1} \propto (T-\theta)^{-1}$ down to 100 K with $\theta$=37 K, and $(T_1T)^{-1}$ exhibits a maximum at 60 K as in the bulk susceptibility. Since Fe 3$d$ spin susceptibility dominates the bulk one, this behavior is expected for a nearly ferromagnetic system from the SCR theory [18]. The details will be reported elsewhere.

We now turn our attention to the low-field region of the isothermal magnetization curves $M(B)$. The data for powder and SPS samples are denoted by solid and open symbols, respectively, in the inset of Fig. 4. For powder samples, $M(B)$ at 5 K shows a hysteresis loop with very small remanent moment of $3.3 \times 10^{-4}$ and $1.0 \times 10^{-3}$ $\mu_B$/Fe for A = Ca and Sr, respectively, values of which are doubled in SPS samples. Although no remanent moment is found for the powder sample with A = Ba, the SPS sample exhibits a similar size of remanent moment as for A = Sr. Chemical inhomogeneity in the SPS samples was carefully examined by EPMA, thereby impurity phases Sb and $FeSb_2$ of approximately 1 vol.% were detected, respectively, on grain boundaries and inside the host grains of 5-20 μm in diameter. In each grain, any compositional gradient larger than the resolution of 1 at.% was not detected. Because both Sb and $FeSb_2$ are diamagnetic [20], they should not be responsible for the observed ferromagnetic behavior. No other Fe-based compounds are known in the ternary phase diagram of A-Fe-Sb to the best of our knowledge.

The very weak ferromagnetism was examined by measuring the temperature dependence of magnetization in various constant fields from 0.01 T to 5 T. The data of $M/B$ taken at $B = 0.01$ T are presented in Fig. 4. For powder samples with A = Ca and Sr, the curves of $M/B$ remarkably increase below 20 and 35 K, respectively. The two sets of data taken in zero-field cooling (ZFC) and field cooling (FC) processes agree with each other. The absence of the upturn in $M/B$ data for the powder sample with A = Ba is consistent with no remanent moment. The onset temperature $T_C$ is raised for SPS samples, and the hysteresis appears between the ZFC and FC data even for A = Ba. The $T_C$ of the SPS samples is tentatively defined as the temperature where the hysteresis sets in. The $T_C$'s for Ca, Sr, and Ba are 54, 48, and 40 K, respectively, above which temperature the disappearance of remanent moment was confirmed. It is noteworthy that the increasing sequence of $T_C$ is that of the decreasing the ionic radius of A ions. This fact hints that the ferromagnetism is related with structural instability. We confirmed that both the remanent moments and the values of $T_C$ are reproduced in several different batches of power and SPS samples, respectively.

If the whole sample undergoes a ferromagnetic transition, then the specific heat should jump at $T_C$ although the jump in an itinerant ferromagnet can be much smaller than in localized magnetic systems [21]. As is shown in Fig. 2, no anomaly in the specific heat is observed. Furthermore, the magnetization $M(B)$ at 2 K increases almost linearly with $B$ up to 5 T, (see the inset (a) of Fig. 3), indicating that the major portion of $M$ originates from a paramagnetic component. These observations indicate that only a portion of the sample undergoes a ferromagnetic transition. Another support for the inhomogeneous and weak ferromagnetism has been obtained by $^{123}$Sb NQR measurements. The NQR signal from a $SrFe_4Sb_{12}$ powder sample slightly broadened by 15% on cooling below $T_C$ = 35 K without showing clear hyperfine splitting.

In order to examine further the weak ferromagnetism in $AFe_4Sb_{12}$, zero and applied longitudinal field muon spin relaxation (μSR) measurements were performed using the pulsed muon source at KEK-MSL. A preliminary analysis of the zero-field spectra has revealed that the initial asymmetry sharply decreases and the relaxation rate strongly increases on cooling below $T_C$ determined by the bulk susceptibility measurement. The differential in the initial asymmetry between 80 K and 1.8 K was used to estimate the volume fraction of the ferromagnetic component. The fractions thus estimated for powder (SPS) samples with A = Ca, Sr, and Ba are 10 (18), 18 (22), 0 (9) %, respectively. Furthermore, the observation of the recovery in the initial asymmetry by applying a longitudinal field of 0.05 T suggests that an internal field of approximately 0.05 T exists in the ferromagnetic part. The results of more detailed analysis will be reported elsewhere.

The above observations lead to the conjecture that the ferromagnetic component arises from either spin polarization around magnetic impurities, as found in Pd(Fe) [22], or some structural deformation in the sample. To understand the larger volume fraction in the SPS sample than in powder sample, we recall that non-uniform distortion in SPS samples was indicated by the broadened the x-ray diffraction peaks. In this regard, it should be noted that the appearance of weak ferromagnetism in the ruthenates with a distorted cubic perovskite structure is very sensitive to both the rotation angle and lattice distortion, as was reported for $Ca_{1-x}Sr_xRuO_3$ and $CaRu_xTi_xO_3$ [3, 23]. Therefore, refinement of the crystal structure of $AFe_4Sb_{12}$ at various temperatures and pressures needs to be done by using high resolution synchrotron x-ray diffraction. It is interesting to note that even without taking lattice distortion into account, recent band structure calculation predicted that the narrow Fe $3d$ band in $AFe_4Sb_{12}$ (A=Ca, Sr, Ba) leads to a weak ferromagnetic ground state [18].

To conclude, we have presented evidence that alkaline-earth metal filled skutterudites $AFe_4Sb_{12}$ (A = Ca, Sr, Ba) are nearly ferromagnetic metals. The magnetic state is found to be intermediate between an itinerant ferromagnetic state in $NaFe_4Sb_{12}$ with a monovalent filler ion and an enhanced paramagnetic state in $LaFe_4Sb_{12}$ with a trivalent filler ion. The spin fluctuations with characteristic temperature of 50 K manifest as the maximum in both magnetic susceptibility and thermopower, as well as the large $\gamma$ value of 100 mJ/mol $K^2$. Furthermore, the emergence of ferromagnetic behavior with a remanent moment of the order of $10^{-3}$ $\mu_B$/Fe was observed below 54, 48, and 40 K for A = Ca, Sr, and Ba, respectively. The volume fraction of the ferromagnetic component was estimated to be 10-20 % by zero-field μSR measurements. Because this volume fraction is beyond the value attributable to impurity effects, we propose that weak ferromagnetism is induced by some sort of distortion of the network of Fe-Sb octahedra. Our observations should stimulate further studies of the role of *d*-band magnetism in the wide variety of physical properties of filled skutterudite compounds.

We thank Y. Shibata for performing electron-probe microanalysis. We are also grateful to K. Takegahara, E. Alleno, P. Haen, and J. Mydosh for useful discussions and sending us preprints prior for publication. This work was financially supported by the COE Research (13CE2002) and the priority area "Skutterudite" (No.15072205) in a Grant-in-Aid from MEXT, Japan.

**Figure captions**

FIG. 1. (a) Temperature dependence of the electrical resistivity $\rho$ and thermopower $S$ for AFe$_4$Sb$_{12}$ (A = Ca, Sr, Ba), LaFe$_4$Sb$_{12}$, and BaRu$_4$Sb$_{12}$. The insets show respectively the $T^2$ dependence of $\rho$ for AFe$_4$Sb$_{12}$ and the x-ray diffraction peak of (763) for power and SPS samples with A=Ba.

FIG. 2. Temperature dependence of the specific heat $C$ for AFe$_4$Sb$_{12}$ (A=Ca, Sr, Ba). The inset shows $T^2$ dependence of $C/T$.

FIG. 3. Temperature dependence of the inverse magnetic susceptibility $B/M$ for AFe$_4$Sb$_{12}$ and LaFe$_4$Sb$_{12}$ in a field of 5 T. Inset (a): magnetization $M$ versus magnetic field $B$ at 2 K, inset (b): $M/B$ versus temperature.

FIG. 4. Temperature dependence of the magnetic susceptibility $M/B$ for AFe$_4$Sb$_{12}$ (A=Ca, Sr, Ba) in a field of 0.01 T for both ZFC and FC processes. Solid and open symbols denote the data for powder and SPS samples, respectively. The inset shows the low-field magnetization curve.

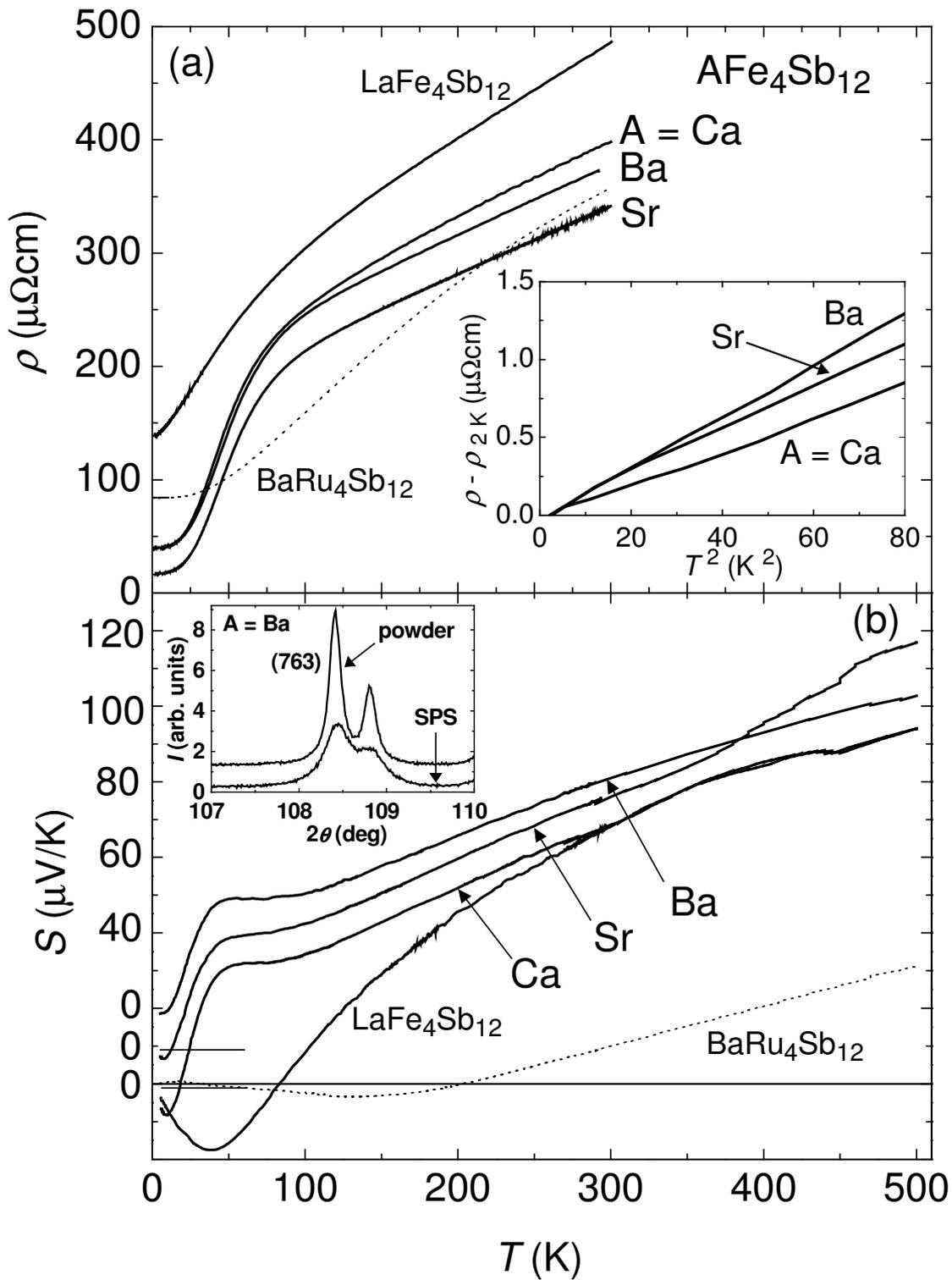

Fig. 1

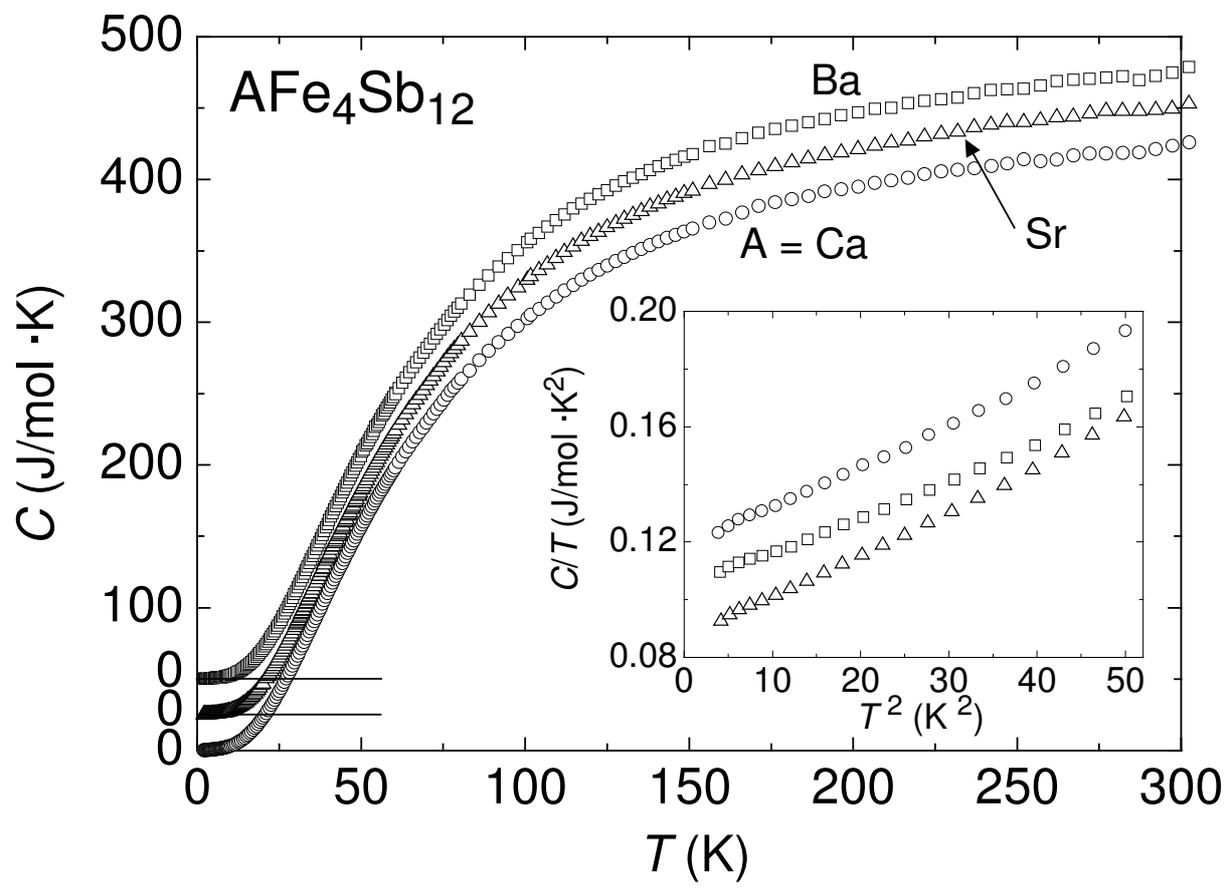

Fig. 2

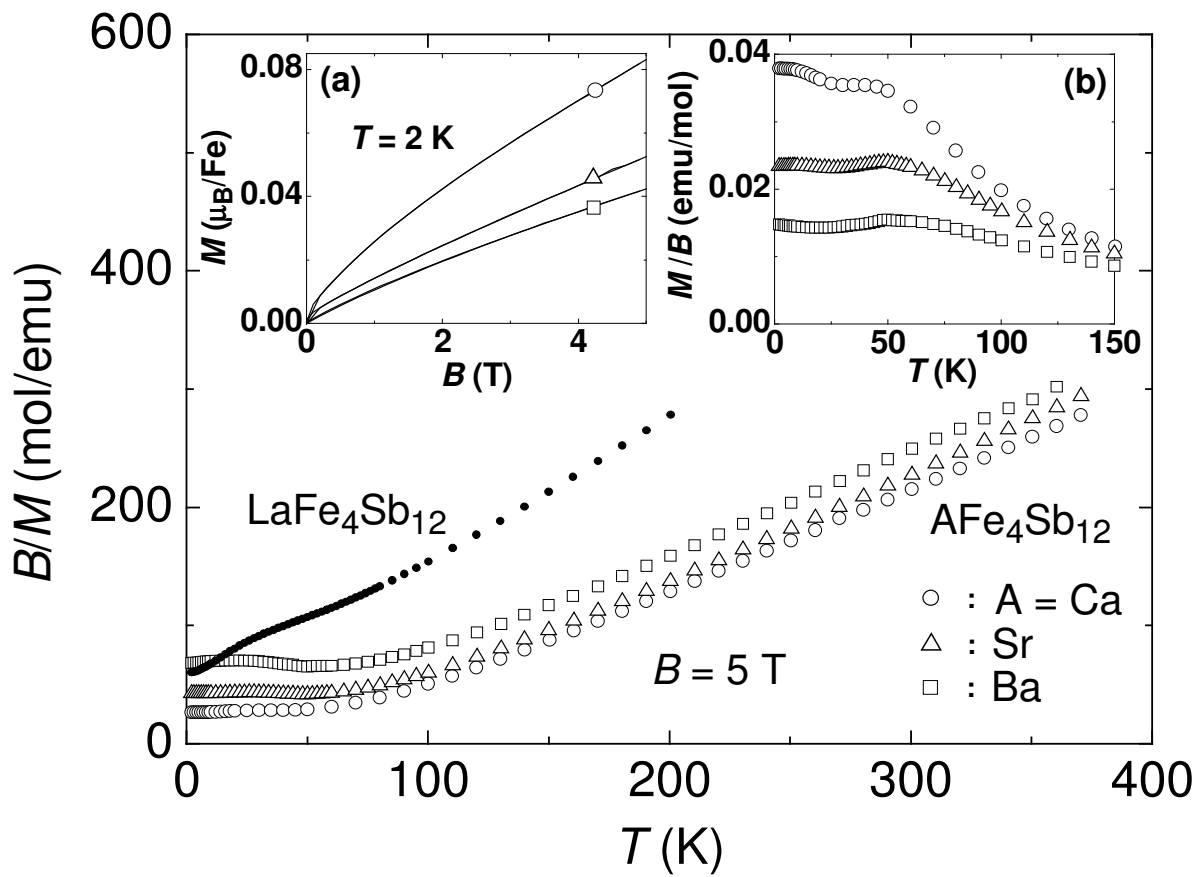

Fig. 3

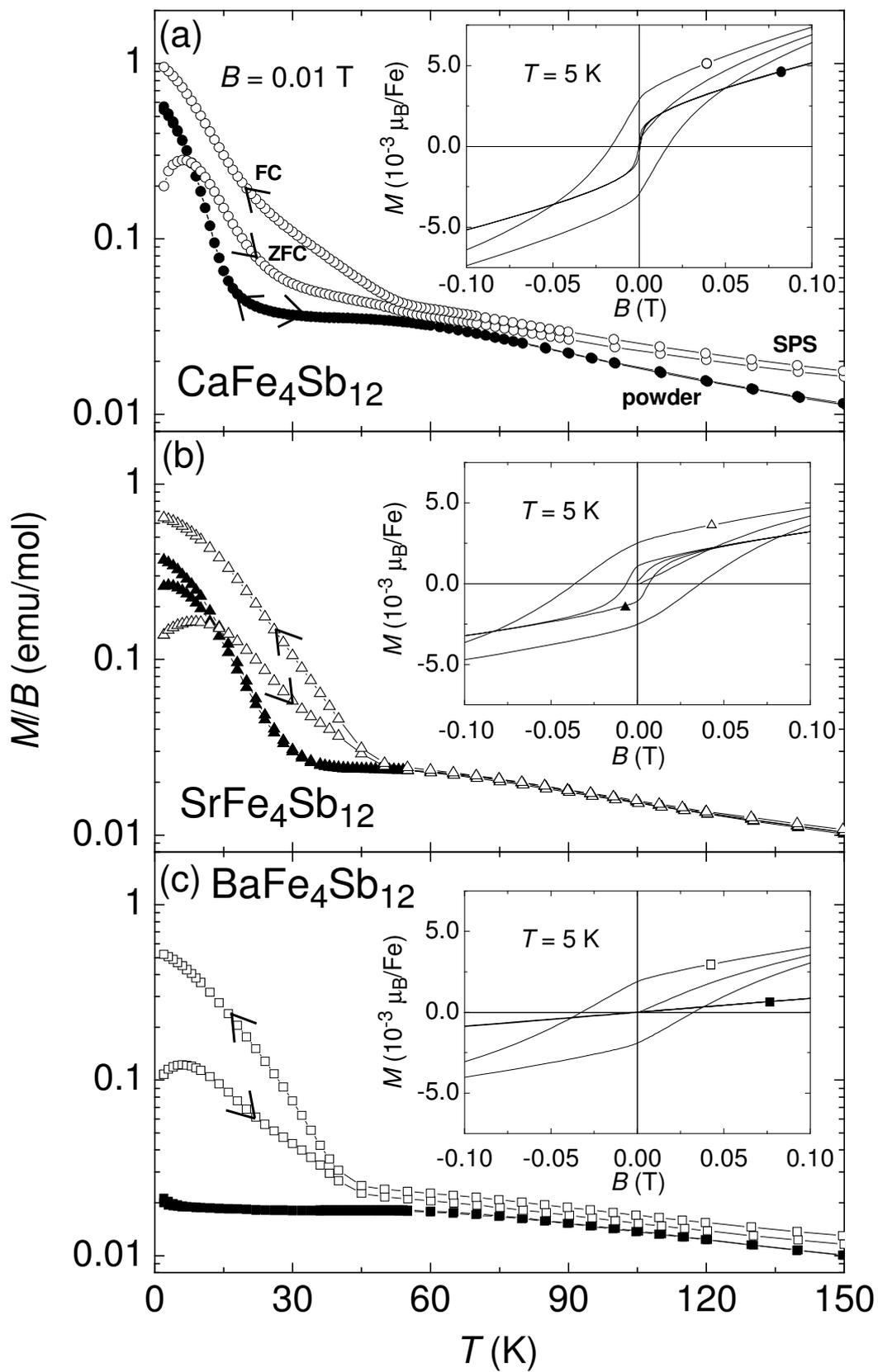

Fig. 4